# Electronic structures across superconductor-insulator transition in Ruddlesden-Popper bilayer nickelate films

Yu Miao[1,2,3†], Runqing Luan[1,2,3†], Yaqi Chen[4,5†], Zhipeng Ou[1,2,3†], Guangdi Zhou[4,5†], Jianchang Shen[1,2,3†], Heng Wang[4,5]*, Haoliang Huang[4,5]*, Xianfeng Wu[4,5], Hongxu Sun[1,2,3], Zikun Feng[1,2,3], Xinru Yong[1,2,3], Yueying Li[4,5], Peng Li[4,5], Lizhi Xu[4,5], Wei Lv[4,5], Zihao Nie[4,5], Changming Yue[4,5], Yu-Jie Sun[4,5], Weiqiang Chen[4,5], Hongtao Yuan[6], Jin-Feng Jia[4,5,7], Qi-Kun Xue[4,5,8]*, Zhuoyu Chen[4,5]*, Junfeng He[1,2,3]*

*[1]Department of Physics, University of Science and Technology of China, Hefei, Anhui 230026, China*

*[2]Hefei National Laboratory, University of Science and Technology of China, Hefei 230088, China*

*[3]Hefei National Research Center for Physical Sciences at the Microscale, University of Science and Technology of China, Hefei, 230026, China*

*[4]State Key Laboratory of Quantum Functional Materials, Department of Physics, and Guangdong Basic Research Center of Excellence for Quantum Science, Southern University of Science and Technology, Shenzhen 518055, China*

*[5]Quantum Science Center of Guangdong-Hong Kong-Macao Greater Bay Area, Shenzhen 518045, China*

*[6]National Laboratory of Solid-State Microstructures, College of Engineering and Applied Sciences, Nanjing University, Nanjing, 210008, China.*

*[7]State Key Laboratory of Micronano Engineering Science, Tsung-Dao Lee Institute & School of Physics and Astronomy, Key Laboratory of Artificial Structures and Quantum Control, Shanghai Jiao Tong University, Shanghai 200240, China*

*[8]Department of Physics, Tsinghua University, Beijing 100084, China*

†These authors contributed equally. *Corresponding author.

Email: jfhe@ustc.edu.cn, chenzhuoyu@sustech.edu.cn, xueqk@sustech.edu.cn, huanghaoliang@quantmsc.cn, wangheng@quantumsc.cn

**High-transition-temperature ($T_C$) superconductivity is recently discovered in Ruddlesden-Popper (RP) nickelate films with extraordinarily strong oxidation [1-6]. While investigating phase diagrams is essential for uncovering the superconducting mechanism, the oxygen-tuned superconductor-insulator transition (SIT) in RP nickelates differs fundamentally from that in cuprates [7] or iron-based systems [8]. Here, we unveil the evolution of electronic structure in RP bilayer nickelate thin films across the SIT, combining angle-resolved photoemission spectroscopy (ARPES) and X-ray absorption spectroscopy (XAS) for both occupied and unoccupied states. In the superconducting state, a coherent quasiparticle band near Fermi level ($E_F$) coexists with an incoherent waterfall feature at high energy, paralleling that in cuprates. Approaching the insulating state with oxygen deficiency, the spectral weight of the occupied coherent quasiparticle band is gradually suppressed, accompanied by pronounced density of states redistribution and orbital reconfiguration in unoccupied states. These results reveal the electronic origin of the SIT in the phase diagram, which transcends carrier doping effects [7-10] and oxygen vacancy states [11,12]. Our findings point to a decisive role of oxygen in shaping the essential electronic landscape of RP bilayer nickelates, offering crucial insights into the superconducting mechanism.**

## Introduction

The recent discovery of superconductivity in nickelates marks an uncharted territory in high-$T_C$ superconductors [1-6,13-23]. Among the superconducting nickelates, square-planar compounds hold a Ni $3d^9$ electron configuration, mimicking the cuprate superconductors [24-28]; Ruddlesden-Popper (RP) compounds, while exhibiting the current record-high $T_C$ in nickelates, are more complex in electronic structure [6,29-34]. On one hand, the nickel site exhibits a ~$3d^{7.5}$ electron configuration, raising a debate on whether the paradigm of doping a Mott insulator is still relevant. On the other hand, insufficient oxygenation of RP bilayer nickelate films drives the system into an insulator [1-6,35,36], although the nature of the insulating phase is still unknown. In this context, revealing the different electronic structures across states with varied oxygen content is crucial for understanding the superconducting mechanism.

Using both laser-based ARPES and synchrotron-based XAS, we have systematically investigated the oxygen-content-tuned electronic structures of RP bilayer nickelate thin films grown by gigantic-oxidative atomic-layer-by-layer epitaxy (GAE) [37,38]. In particular, an ultra-high vacuum (UHV) cryogenic sample transfer technique [32,34] was implemented to ensure the preservation of surface oxygen content for ARPES measurements.

## Transport phase diagram

Superconductivity in RP bilayer thin films is highly sensitive to oxygen content [1-6,35,36]. *In situ* fine tuning oxygen content (Fig. 1a) within the transport measurement cryostat (as detailed in Methods) reveals a SIT in resistivity-temperature ($R$-$T$) curves (Fig. 1b). At sufficiently high oxygen content, the films exhibit zero-resistance superconductivity at low temperatures, though $T_C$ is slightly suppressed in the most oxygen-rich regime [39]. With decreasing oxygen content, the normal state resistivity (at 200 K) increases and the superconducting state is progressively degraded. Yet, a resistive drop indicative of Cooper pair formation remains discernible. When the oxygen content is further reduced, the thin film becomes insulating at low temperature, and the $R$-$T$ curves exhibit metal-insulator transitions shifting progressively towards higher temperatures. In the heavily oxygen-deficient limit, an insulating behavior is observed across the entire temperature range. We quantify the oxygen content via an empirical relation $p = 1031/R$ (200 K) with the oxygen content of the initial state set to 1, following prior conventions [40,41]. As such, a color-mapped temperature-oxygen content phase diagram is drawn to illustrate the transition between superconducting and insulating regimes (Fig. 1c). Superconductivity remains robust when $0.3 \leqslant p \leqslant 1.0$, with $T_C^{Onset}$ and $T_C^{50\%}$ plateauing around 50 K and 38 K, respectively; and the SIT starts to appear in the oxygen-deficient regime (e.g. $p<0.3$).

## Electronic structure of the superconducting state

Laser-based ARPES measurements are first carried out in the superconducting state (Fig. 2, also see Extended Data Fig. 1). Photoelectron intensity plot along the (0,0)-(π,π) direction is shown in a wide energy range (Fig. 2a). A dispersive energy band is identified near $E_F$, which is connected by a vertical waterfall-like spectrum at high energy (Fig. 2a).

The band dispersion is quantitatively extracted by fitting momentum distribution curves (MDCs, Fig. 2b, also see Extended Data Fig. 2), and a nearly vertical dispersion is indeed observed below -0.2 eV (Figs. 2b,e), resembling the characteristic waterfall feature reported in cuprates [10]. It is important to note that a substantial broadening of MDC peaks is also observed in the vertical dispersion region (Fig. 2b), demonstrating an incoherent nature of this high-energy waterfall. Energy distribution curves (EDCs) provide another approach to extract band dispersion from a two-dimensional photoemission spectrum [10,42]. In principle, the EDC- and MDC-derived band dispersions should be identical in a weakly interacting Fermi liquid system. However, the energy band derived by EDC peaks of the superconducting film exhibits substantial differences from that extracted from the MDCs (Fig. 2e). The dispersive band near $E_F$ is well captured by EDCs, exhibiting both superconducting coherence peaks and dispersion kink for electron-boson coupling (Figs. 2c-e), but the high-energy waterfall is completely absent in the EDC-derived dispersion (Fig. 2e, also see Extended Data Fig. 3 for the similar effect along an off-diagonal direction of the Brillouin zone). Such a discrepancy between the EDC- and MDC-derived dispersions has also been reported in cuprates with high-energy waterfall [42]. This experimental observation can be phenomenologically reproduced by simulating the coexistence of a quasiparticle band at low energy and incoherent waterfall-like spectral weight at high energy (Fig. 2f-i).

**Electronic structure across the superconductor-insulator transition (SIT)**

Evolution of energy bands across the SIT is investigated by *in situ* tuning the oxygen content of nickelate films in laser-based ARPES system (Fig. 3, see Methods). As the oxygen content decreases, instead of a carrier doping induced band shift, a progressive spectral weight suppression of the low-energy quasiparticle band is observed (Fig. 3a). Such an evolution is quantified by analyzing the spectral weight in three steps: first, all spectra are integrated over momentum, forming momentum-integrated EDCs (Fig. 3b); second, a fitted line of the momentum-integrated EDC with the lowest oxygen content is subtracted from all momentum-integrated EDCs to isolate the effect of spectral weight suppression (Fig. 3c); third, the suppressed spectral weight is quantified and shown as a function of oxygen loss (Fig. 3d). As illustrated by the schematic in Fig. 3f, the quasiparticle spectral weight near $E_F$ gradually fades out with reducing oxygen content, while the high-energy spectral weight still persists. Such an evolution of the electronic structure echoes our transport observations, where the lowered density of electrons near $E_F$ and the loss of quasiparticle coherence may collectively drive the SIT.

The unoccupied states also exhibit substantial changes across SIT, as examined by XAS measurements. First, a pre-peak is apparently observed at energies prior to the oxygen *K* edge, signifying the hybridization between O 2*p* states and Ni 3*d* states near $E_F$. As SIT takes place, the integrated area of the O *K* edge pre-peak (Fig. 4a, inset) progressively diminishes, indicating suppressed hybridization between the electronic states near $E_F$. This is aligned with the spectral weight suppression near $E_F$ observed in ARPES (Fig. 3). Beyond the pre-peak, the higher-energy peaks at ~533 eV and ~536 eV reflect the hybridization of O 2*p* states with extended Ni 4*sp* and rare-earth (e.g. La) 5*d* states. These two peaks are also reshaped with varied oxygen content (Fig. 4a), highlighting an overall redistribution of the electronic states.

X-ray linear dichroism (XLD) analysis at the Ni $L_2$ edge further reveals orbital contrast between superconducting (Fig. 4b) and insulating (Fig. 4c) states (also see Extended Data Fig. 4). In the Ni-O octahedral coordination, the out-of-plane $d_{z^2}$ orbital couples selectively to $I_c$ (out-of-plane polarization), while the in-plane $d_{x^2-y^2}$ dominates $I_{ab}$ (in-plane polarization). In the superconducting state (Fig. 4b), the finite magnitude of both $I_{ab}$ and $I_c$ indicates the coexistence of $d_{x^2-y^2}$ and $d_{z^2}$ states near $E_F$. This is consistent with previous ARPES reports of superconducting samples, in which a $d_{z^2}$ dominant pocket surrounds the Brillouin zone corner [6,30,32,34]. $I_{ab}$ significantly exceeds $I_c$, reflecting pronounced orbital polarization with a dominant unoccupied $d_{x^2-y^2}$ character above $E_F$. Upon transitioning to the insulating state, this orbital contrast markedly diminishes as the difference between $I_{ab}$ and $I_c$ decreases (Fig. 4c). This orbital reconfiguration echoing our ARPES observations, defies a simple rigid band shift picture induced by carrier doping.

## Discussion

We now discuss the implications of our observations. First, the coexistence of a coherent low-energy quasiparticle band and an incoherent high-energy waterfall-like feature observed in the superconducting state resembles those of cuprates. Such a behavior cannot be explained by simple band renormalization or high-energy mode coupling (Extended Data Fig. 5), nor can it be attributed to matrix element effects associated specifically to copper oxides [43]. To identify the origin of the incoherence feature, we further consider oxygen vacancy states and polaronic states. While oxygen vacancy states may produce incoherent high energy features which tend to be intensified with increasing oxygen vacancies [11,12], the nickelate high-energy waterfall appears in the superconducting state with sufficient oxygenation, exhibiting little response to the oxygen content change. Similarly, although polaronic states may also produce incoherent phonon satellites [44], the hallmark replica bands of polaronic shake-off excitations are absent in our results. Therefore, the observed electronic structures are more likely associated with electronic correlation effects [45,46]. Within the framework of a doped Mott insulator, a quasiparticle band can split off from the Hubbard band with decreasing Coulomb interaction, forming an incoherent waterfall connecting both bands [46]. In this regard, it would be interesting to examine whether Hubbard model is relevant to the $\sim 3d^{7.5}$ electron configuration in RP bilayer nickelates. It would also be encouraging to explore whether more itinerant models with electronic correlation can generate these features. Regardless of the particular theoretical model, our results demonstrate that the synergy of quasiparticles and electronic correlation is indispensable for high-$T_C$ nickelate superconductivity.

Second, the SIT in the RP bilayer nickelate films is found to be driven by a gradual suppression of the coherent quasiparticle band. Notably, while oxygen loss is generally expected to induce electron doping, the observed electronic evolution is distinct from the typical rigid band shift. In oxide perovskites, oxygen vacancy states are also frequently reported, producing both localized high-energy states and itinerant electrons near $E_F$ [11,12]. However, such new states would not suppress the original coherent quasiparticle band. Disorder effect can appear under sufficiently strong oxygen reduction, which might suppress itinerant electronic states. However, the concurrent density of states redistribution and orbital reconfiguration observed in the unoccupied states defy a simple disorder effect. In this sense, the concomitant changes of occupied and unoccupied states near $E_F$, alongside an almost unchanged high-energy incoherent band and a macroscopic SIT, indicate that oxygen stoichiometry in RP nickelates plays more critical roles than tuning carriers, inducing vacancy states, or simple disorder effect: it may fundamentally reshape the electronic landscape near $E_F$.

Third, the spectral weight suppression of the $d_{x^2-y^2}$-derived band hints at a potential loss of in-plane oxygen atoms. This in-plane oxygen deficiency could exert a dual effect on the system's electronic properties: it likely reduces the density of states available for Cooper pairing and simultaneously disrupts in-plane coupling, which is critical for establishing the global phase coherence required for superconductivity. In this context, it would be interesting to examine whether local Cooper pairs can exist in the insulating region with lowered electron density of states near $E_F$.

Fourth, our XLD data provides tentative support for the scenario of in-plane oxygen vacancy (Fig. 4d). Under the synergistic effect of strain and Jahn-Teller distortion, the $e_g$ orbitals split with the average $d_{x^2-y^2}$ state energy lying above the average $d_{z^2}$ state energy due to crystal field. In principle, apical oxygen vacancies would lower the $d_{z^2}$ energy and increase this crystal field splitting, whereas in-plane vacancies would lower the $d_{x^2-y^2}$ energy and reduce the splitting. The observed reduction in orbital contrast between $I_{ab}$ and $I_c$ in the insulating state is more consistent with a decreased energy difference between the $e_g$ orbitals, lending weight to the dominance of in-plane oxygen loss. Our data do not preclude the concurrent presence of apical vacancies [47], but the spectroscopic implication of dominant in-plane oxygen loss aligns with the recent observations by scanning transmission electron microscopy [4].

In summary, we have systematically mapped the electronic structure evolution across the SIT in RP bilayer nickelate films using ARPES and XAS. The observed electronic structure in superconducting state, comprising a coherent quasiparticle band and a high-energy incoherent waterfall, parallels the cuprates and highlights electronic correlation as a fundamental signature of the superconducting state. The simultaneous suppression of quasiparticle spectral weight and the redistribution of unoccupied states reveal an unconventional SIT that defies a simple carrier-doping description. These results establish the decisive role of oxygen stoichiometry in engineering the essential electronic landscape required for high-$T_C$ superconductivity in RP bilayer nickelates.

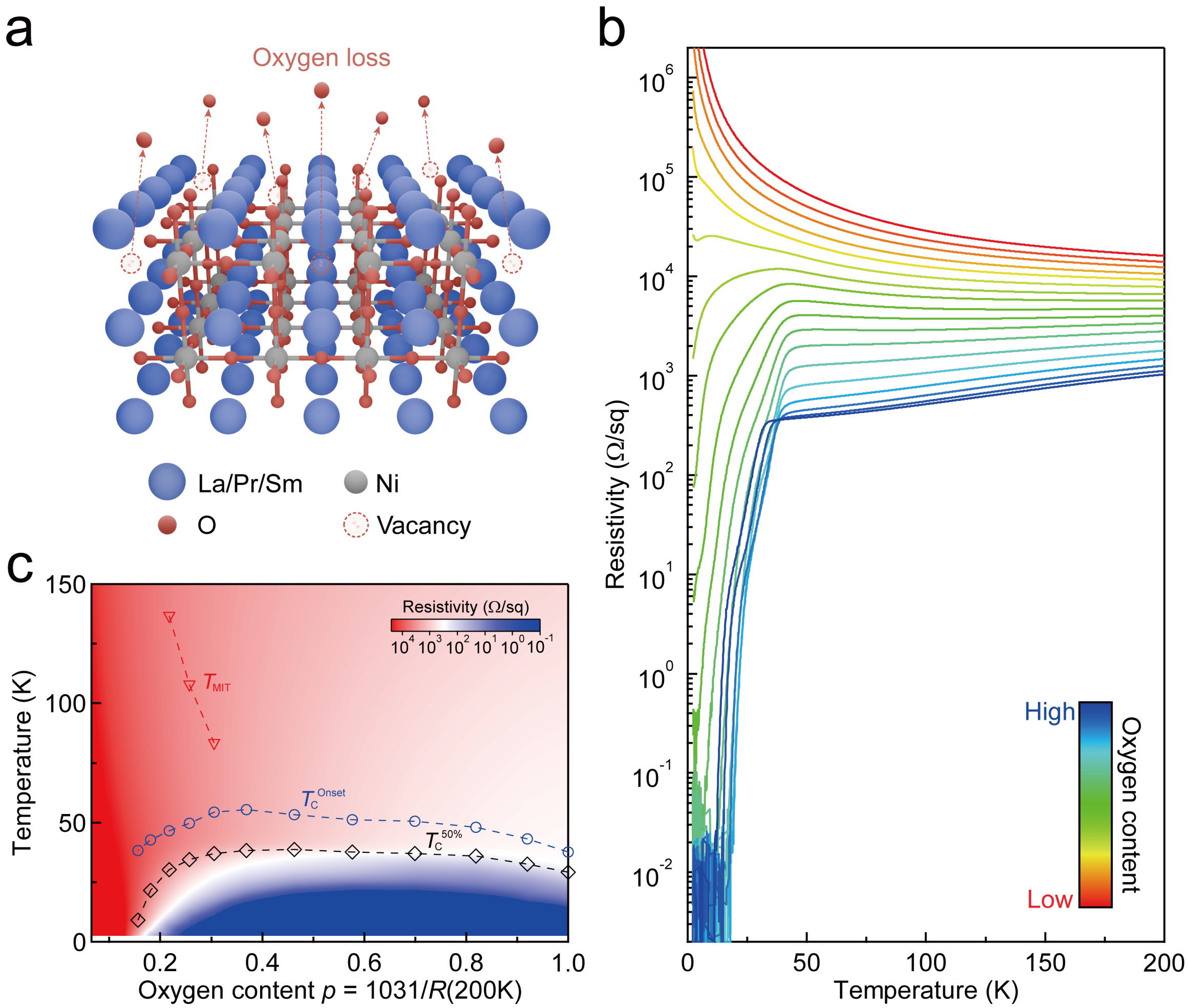

**Fig. 1. Oxygen-tuned superconductor to insulator transition.** (**a**) Schematic of oxygen loss in nickelate films. (**b**) Temperature dependence of the resistivity curve, measured on a 3-unit-cell thick $La_{2.31}Pr_{0.24}Sm_{0.45}Ni_2O_7/SrLaAlO_4$ sample with different oxygen content $p$. (**c**) Color map of resistivity as a function of temperature and oxygen content $p$. The oxygen content is quantified by using the empirical relation $p = 1031/R(200\text{K})$. The extracted $T_{\text{MIT}}$, $T_{\text{C}}^{\text{Onset}}$ and $T_{\text{C}}^{50\%}$ are overlaid. $T_{\text{C}}^{50\%}$ is defined by the temperature at which the resistivity drops to half of the value at $T_{\text{C}}^{\text{Onset}}$.

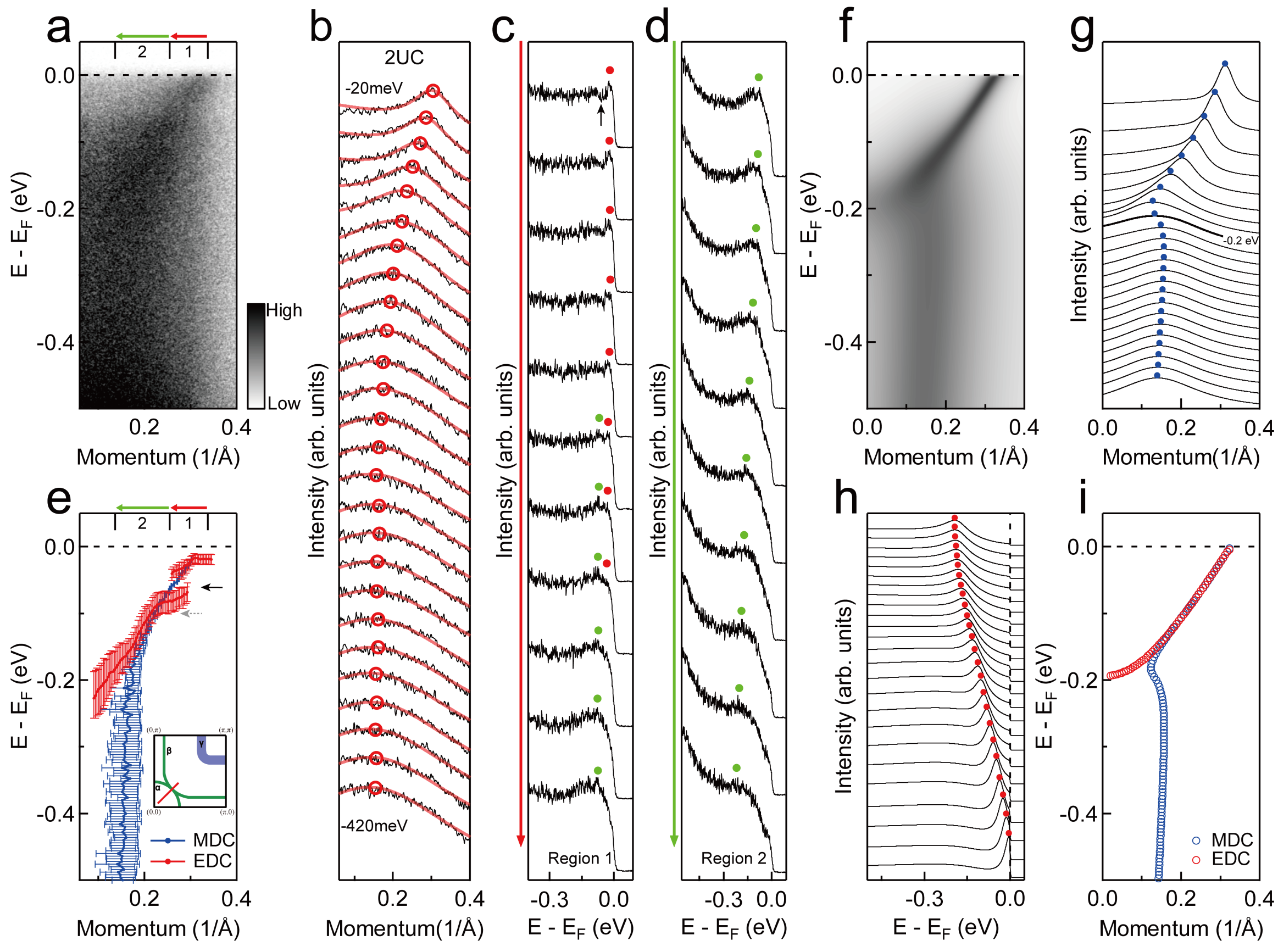

**Fig. 2. Comparison of EDC- and MDC-derived band dispersion along the (0,0)-(π,π) direction.** (**a**) Photoelectron intensity plot as a function of momentum and energy, measured on a superconducting 2-unit-cell thick $La_{2.31}Pr_{0.24}Sm_{0.45}Ni_2O_7/SrLaAlO_4$ thin film at 10 K. An integrated EDC in the background region (Extended Data Fig. 3) is taken as the spectral background and subtracted from the raw image to better visualize the dispersion. Note that the background subtraction would not change the MDC-derived band dispersion (Extended Data Fig. 3). (**b**) MDCs extracted from panel (a). Circles mark the MDC peak positions, and the red curves are fits overlaid on the data. A Lorentzian function with a linear background is used for the fitting. (**c, d**) EDCs in momentum regions 1 and 2, respectively. The momentum regions are indicated by red and green arrows in (a), respectively. EDC peaks are marked by circles. Black arrow in (c) indicates the EDC dip feature of the reported ~70 meV energy mode [32]. (**e**) Comparison between the MDC-derived (blue circles) and EDC-derived (red circles) band dispersion for the same measurement. Error bars indicate uncertainties in the extraction of the band dispersions. The black arrow marks the energy feature at ~70 meV, and the gray dashed arrow indicates another possible energy feature at ~100 meV. (**f**) Simulated spectral function $A(k,\omega)$ with the coexistence of a low-energy quasiparticle band and a high-energy incoherent band. (**g**) Raw MDCs in (f). MDC peaks are marked by blue circles. (**h**) Raw EDCs in (f). EDC peaks are marked by red circles. (**i**) Band dispersion extracted from MDC (blue circles) and EDC (red circles) peaks, respectively.

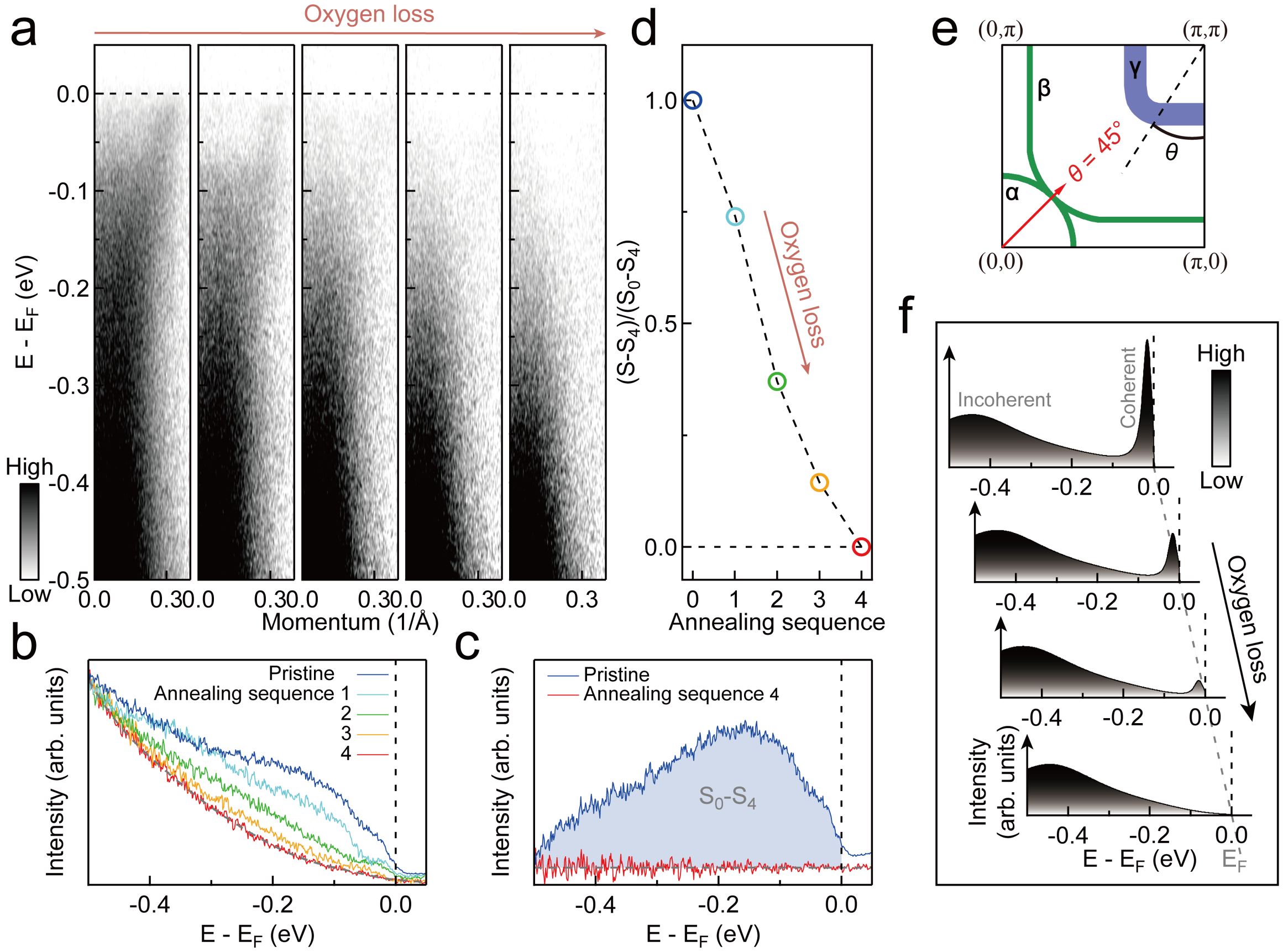

**Fig. 3. Evolution of the electronic structure as a function of oxygen content.** (**a**) Photoelectron intensity plot of the band structure of a 1-unit-cell thick $La_{2.85}Pr_{0.15}Ni_2O_7/SrLaAlO_4$ film, measured along (0,0)-(π,π) at 10 K. The oxygen content is gradually reduced by *in situ* annealing (from left to right). (**b**) Momentum-integrated EDCs from (a). The EDCs are labeled by annealing sequences. The dashed line represents a polynomial fit to the momentum-integrated EDC at annealing sequence 4. (**c**) Representative momentum-integrated EDCs, subtracted by the fitting curve of the momentum-integrated EDC at annealing sequence 4. The shaded area quantifies the spectral weight difference between pristine sample ($S_0$) and annealing sequence 4 ($S_4$). (**d**) Spectral weight suppression as a function of annealing sequence, normalized by the equation $(S\text{-}S_4)/(S_0\text{-}S_4)$. (**e**) Location of the momentum cut in the Brillouin zone, marked by the red arrow. (**f**) Schematic evolution of the electronic structure as a function of oxygen content. The quasiparticle peak is suppressed with oxygen loss.

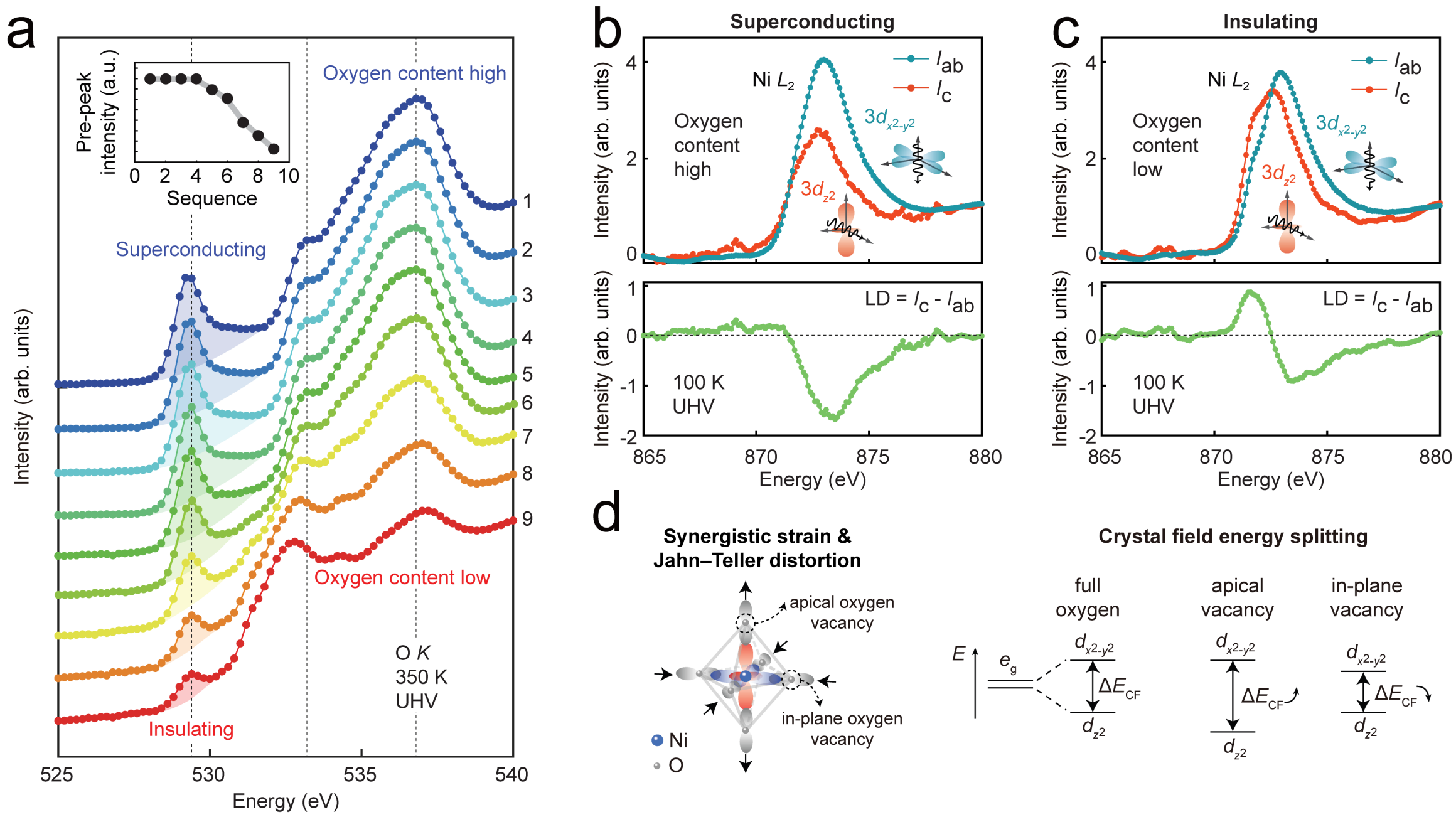

**Fig. 4. Evolution of unoccupied electronic states as a function of oxygen content.** (**a**) XAS of the O *K* edge measured on a 3-unit-cell thick $La_{2.85}Pr_{0.15}Ni_2O_7/SrLaAlO_4$ film during continuous oxygen loss in ultra-high vacuum (UHV) at 350 K. Inset: The pre-peak intensity area integral decreases as the superconducting state with high oxygen content transitions to the insulating state with low oxygen content. (**b,c**) Polarization dependent XAS of Ni $L_2$ edge and linear dichroism (LD) signal in the superconducting state and the insulating state, respectively. Measurement temperature is 100 K under UHV. (**d**) Schematic of oxygen vacancy induced orbital reconfiguration in the unoccupied states.